\documentclass[aps,preprint,showpacs,floatfix,twocolumn,10pt]{revtex4-1}
\usepackage{graphicx}
\begin{document}
\title{Ensemble Equivalence for Counterion Condensation on a Two
Dimensional Charged Disc}
\author{Anoop Varghese}
\email{anoop@imsc.res.in}
\author{Satyavani Vemparala}
\email{vani@imsc.res.in}
\author{R. Rajesh}
\email{rrajesh@imsc.res.in}
\affiliation{The Institute of Mathematical Sciences, C.I.T. Campus,
Taramani, Chennai 600113, India}
\begin{abstract}
We study the counterion condensation on a two dimensional charged disc in the
limit of infinite dilution, and compare the energy--temperature 
relation obtained from the canonical free energy and microcanonical 
entropy.  The microcanonical entropy is piecewise linear in energy, 
and is shown to be concave for all energies. As a result, even though
the interactions are long-ranged, the energy--temperature relation and hence the counterion
condensation transition points are identical in both the ensembles.
\end{abstract}
\date{\today}
\maketitle

Polyelectrolytes are polymers that release
counterions into a polar solvent, making the polymer backbone 
charged \cite{dob,netz1,yan}. They are ubiquitous in biological systems --
examples include DNA~\cite{bloom}, F-actin, microtubles and
tobacco mosaic virus~\cite{tang}. 
On increasing the linear charge density of the polymer
backbone beyond a threshold value, 
counterions 
start condensing onto the polymer~\cite{manning,muthukumar}. 
The interactions due to the condensed counterions lead to complex phase
diagrams~\cite{dob,netz1,pincus,jaya}, and phenomena such as 
collapse of an extended polyelectrolyte chain~\cite{wink,brilliantov,varghese2011}, 
and aggregation of similarly charged 
polyelectrolyte chains~\cite{bloom,stevens,holm}. This counterion
condensation transition (CCT) has been
mostly studied in the mean field limit, 
for idealised systems like cylinder
\cite{manning,netz2,netz3}, planes~\cite{netz4,manning2,manning3} and
spheres~\cite{manning2}, due to the analytical intractability of real
polyelectrolyte systems.
In recent work, the partition function of a two dimensional system of
counterions around
a charged disc was evaluated in the limit of infinite 
dilution \cite{burak}.  
The theoretically obtained critical points of CCT and the 
dependence of energy on temperature matched with 
results from Monte Carlo simulations~\cite{netz2,netz3}. 

Electrostatic interactions are long-ranged, i.e., 
in $d$-dimensions, they decay slower than $r^{-d}$ at large distances $r$. 
A possible consequence of long-range interactions and non-additivity of
energy is the inequivalence of
different statistical ensembles \cite{campa,ruffobook}. This
inequivalence may be manifested as negative specific heat or 
magnetic susceptibility in the
microcanonical ensemble, different order of phase transitions, and
different critical points~\cite{campa,ruffobook,barre}. As systems that show CCT
are long-ranged, it would be of interest to know if CCT is the same in
different ensembles.
However, almost all such models are only studied in the
canonical ensemble. 
In this paper, we solve the model of CCT in Ref.~\cite{burak} of a system
of counterions condensing onto an oppositely charged disc in
two dimensions in the microcanonical ensemble to address the question of
ensemble equivalence of CCT. We first obtain the expression for
entropy and show that it is piecewise linear in energy. This 
entropy is then 
shown to be concave everywhere in energy and volume. The CCT 
temperature and the
energy-temperature relations obtained in the microcanonical ensemble
are shown to be identical to those  obtained from
the calculation in the canonical ensemble \cite{burak}. 
Thus CCT, at least in this solvable case, is shown to be equivalent in the 
canonical and microcanonical ensembles, despite the interactions being long-ranged.

Examples of non-additive long-ranged 
systems that show ensemble inequivalence include spin 
systems~\cite{mukamel,ellis3}, self-gravitating systems~\cite{padmanabhan,miller,stahl}, 
plasma~\cite{kiessling,smith}, two dimensional hydrodynamic 
systems~\cite{ellis1,ellis2} and other 
model systems~\cite{thirring,antoni,mukamel2}. In 
the last decade, there have been various attempts to compare the 
solutions of long-ranged systems obtained in the
microcanonical and canonical ensembles~\cite{campa}. 
However, the number of exactly solvable models where one can pinpoint the
equivalence or inequivalence are limited, since the calculation of 
microcanonical entropy is difficult even for systems with short-ranged
interactions.  The solution presented in this paper will add to this list of
exactly solvable models.

Consider a uniformly charged disc of charge $q$ and radius $a$. 
$N$ counterions, each carrying a charge $-q'$, are
distributed in the annular region 
between the charged disc and a circular boundary of radius $R$. 
Overall charge neutrality is achieved by choosing $q=Nq'$. 
Let ${\bf r}_i$ be the position of counterion
$i$ in a coordinate system with origin at the center of the disc. 
The Hamiltonian of the system is
\begin{equation}
H= 2\chi \sum_{i=1}^{N}\ln\left(\frac{r_{i}}{a}\right)-
\frac{\chi}{N}\sum_{i \neq
j} \ln\left(\frac{r_{ij}}{a}\right),
\label{eq1}
\end{equation}
where $r_{i} = |{\bf r}_i|$ and
and $r_{ij}= | {\bf r}_i - {\bf r}_j|$.
The parameter $\chi=qq'=Nq'^{2}$ measures the strength of the 
electrostatic interaction.
The permittivity $\epsilon$ has been set equal to 
$1/(4\pi)$. The thermodynamics of this system was studied in the
canonical ensemble in Ref.~\cite{burak}. We now obtain the expression
for the partition function following closely the steps in
Ref.~\cite{burak}. 

The Hamiltonian can be rewritten in terms of
$u_{i}=\ln(r_{i}/a)$ as
\begin{eqnarray}
H&=&\chi \left(1+\frac{1}{N}\right)\sum_{i=1}^{N}u_{i} \nonumber\\
&-&\frac{\chi}{2N}\sum_{i\neq
j}\ln\left[2\cosh(u_{i}-u_{j})-2\cos\theta_{ij}\right],
\label{eq3}
\end{eqnarray}
where $\cos\theta_{ij}= {\bf r_i} \cdot{\bf r_j}/(r_i r_j)$. 
This Hamiltonian is analytically intractable. However, some simplifications
occur when the limit of infinite dilution, $N/R^2 \rightarrow 0$, is considered.
This corresponds to the limit of finite number of counterions in infinite
volume, similar to the limit considered in Manning condensation \cite{manning}.
In this limit, counterions are far away from each other. 
Thus,  $|u_{i}-u_{j}|\gg1$, 
and the summand in the second term of
Eq.~(\ref{eq3}) may be approximated by $|u_{i}-u_{j}|$,
since $2\cosh(u_{i}-u_{j})\approx \exp\left(|u_{i}-u_{j}|\right)$. 
In this limit, the partition 
function of the system was calculated in Ref.~\cite{burak} and 
is given by (in slightly different notation),
\begin{eqnarray}
Z&=&\sum_{k=0}^{N}\prod_{\begin{array}{ll}m=0\\m\neq
k\end{array}}^{N}
\left[\frac{\xi\left(k,m\right)}{2\left(k-m\right)\chi\left[\beta-
\frac{\xi\left(k,m\right)}{\chi}\right]}\right] \nonumber\\
&\times&
\exp\left[\frac{-2\left(N-k\right)\chi
L}{\xi\left(N,k\right)}\left(\beta-\frac{\xi\left(N,k\right)}{\chi}\right)
\right],
\label{eq14}
\end{eqnarray}
where $L=\ln(R/a)$, $\beta$ is the inverse temperature and 
\begin{equation}
\xi\left(k,m\right)=\frac{2N}{2N-k-m+1}.
\label{eq13}
\end{equation}

In the limit $L\rightarrow\infty$ keeping $N$ fixed, 
the sum in Eq.~(\ref{eq14}) is dominated by the largest summand.
Given a value of $\beta \chi$, the $k$ corresponding to the largest
summand is obtained by solving $\beta\chi=N/(N-k+1)$
for $k$ and taking the integer part~\cite{burak}. As $\beta \chi$ is increased, $k$
changes by unity, resulting in non-analytic behaviour of the
free energy. These critical points occur at
\begin{equation}
(\beta\chi)_c=\frac{N}{N-k+1}=\xi\left(k-1,k\right), ~k=1,2,\ldots N.
\label{eq21}
\end{equation}
The first transition is at $\beta\chi=1$ $(k=1)$, corresponding to the Manning
condensation temperature on cylinders~\cite{manning}.

To show the ensemble equivalence, we now compute the entropy
in the microcanonical ensemble from the canonical partition 
function in Eq.~(\ref{eq14}). We then show that the entropy is concave
everywhere and the CCT temperatures coincide in both the ensembles.

The density of states $g(E)$ is defined as
\begin{equation}
g\left(E\right)=\frac{1}{N!}\int\prod_{i=1}^{N}dr_{i}r_{i}
d\theta_{i}~\delta(E-H).
\label{eq16}
\end{equation}
$g(E)$ is obtained from the partition function $Z$ by
performing an inverse Laplace transform, i.e., 
\begin{equation}
g(E)=\frac{1}{2\pi
i}\int_{-i\infty+c}^{+i\infty+c}d\beta~Z(\beta)~\exp(\beta E),
\label{eq18}
\end{equation}
where the constant $c$ is to be chosen such that the path of integration lies
to the right of all the poles of $Z(\beta)$. 
This corresponds to $\xi(k,m)/\chi<c$ for all 
$k$ and $m$.  We evaluate the above integral by the method of residues,
by closing the contour of integration with a semi-circle that is either to
the left or right such that the contribution to the integral from the
semi-circle is zero. If $E$ and $k$ do not satisfy the condition,
\begin{equation}
E-\frac{2\left(N-k\right)\chi L}{\xi\left(k,N\right)}>0,
\label{eq20}
\end{equation}
then the contour of integration is closed to the right in the 
complex plane. The closed contour does not enclose any of
the the poles of the partition function, and, hence, the contribution
to $g(E)$ from such $k$ is zero. On the other hand, if $E$ and $k$ satisfy
the condition in Eq~(\ref{eq20}), the contour is closed to the left, 
and $g(E)$ is the sum of the residues of the partition function, and is 
given by 
\begin{widetext}
\begin{equation}
g(E)=\sum_{k=k^*}^{N} 
\sum_{\begin{array}{ll}l=0\\l\neq k\end{array}}^{k^*-1}
\frac{\xi\left(k,l\right)}{2\left(k-l\right)\chi}
\exp\left[\frac{\left(N-k\right)\left(N-l\right)\xi\left(k,l\right)L}{N}
+\frac{\xi\left(k,l\right)E}{\chi}\right]
\prod_{\begin{array}{lll} m=0\\ m\neq k \\ m\neq l\end{array}}^{N}
\left[\frac{N}{\left(k-m\right)\left(l-m\right)\xi\left(k,l\right)}\right],
\label{eq19}
\end{equation}
\end{widetext}
where $k^{*}$ is the smallest value of $k$, given $E$, that satisfies
Eq.~(\ref{eq20}). In writing Eq.~(\ref{eq19}), we have used the relation
\begin{equation}
\xi\left(l,k\right)-\xi\left(m,k\right)=
\frac{\left(l-m\right)\xi\left(l,k\right)\xi\left(m,k\right)}{2N},
\end{equation}
derived from Eq.~(\ref{eq13}). The terms in the summation with 
$l\geq k^*$ do not contribute to $g(E)$ because the summand is 
antisymmetric in $k$ and $l$. This leads to an upper bound $k^*-1$ for $l$.
Also, the maximum possible value of the energy of the system is $(N+1)\chi L$, 
corresponding to $u_i=L$ for all $i$. It then follows from 
Eq.~(\ref{eq20}) that the lowest value of $k^*$ is $1$.

The density of states $g(E)$ in Eq~(\ref{eq19}) is a sum of exponentials of
the form $\exp[f(k,l,E/L) L]$, where the function $f$ is derivable from
Eq~(\ref{eq19}). In the limit $L \rightarrow \infty$, keeping $N$ fixed, 
the sum is dominated by that $k,l$ which maximize the function $f$.
$E$ is implicitly dependent on $L$ and we
make this dependence explicit  by converting 
the inequality in Eq.~(\ref{eq20}) into an equality by
replacing $k$ by $k^*-\delta$, where $0<\delta<1$.
Thus, eliminating $E$, $f$ becomes a function of
$k$, $l$ and $k^*$. It is then easy to show  that $f$
increases with $l$ for $k\geq k^*$, and decreases with $k$ for $l\leq k^*$. 
As a result, the dominant contribution to $g(E)$
comes from the term with $k=k^*$ and $l=k^*-1$.

\begin{figure} 
\includegraphics[width=\columnwidth]{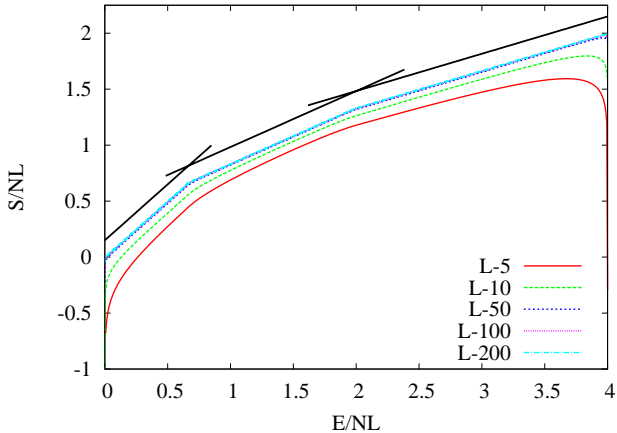}
\caption{The variation of microcanonical entropy $S$ with energy $E$ is shown
for different $L$. The data are for $N=3$ and $\chi=3$. 
The solid black lines correspond to the limiting curve obtained from 
Eq.~(\ref{eq23}). They have been shifted upwards and extended to the left and
right for clarity since the unshifted lines are 
indistinguishable from the curve for $L=200$.} 
\label{fig1}
\end{figure}

By considering only the largest term, we obtain the entropy $S=\ln[g(E)]$ 
to be
\begin{equation}
\lim_{L\rightarrow \infty} \frac{S}{N L}=1-\frac{k^*}{N}+
\frac{\xi(k^*,k^*-1)}{\chi} \frac{E}{N L},
\label{eq23}
\end{equation}
where $\xi(k^*,k^*-1)= N/(N-k^*+1)$. In Fig.~\ref{fig1}, we show the
dependence of entropy on energy, obtained by  considering all the terms in 
$g(E)$ [see Eq.~(\ref{eq19})], and compare it with the entropy in
Eq.~(\ref{eq23}). On increasing $L$, $S/(NL)$ approaches the limiting curve
with $N$ linear portions, determined by Eq.~(\ref{eq23}).

We now show that the entropy is concave in energy $E$.  In
Eq.~(\ref{eq23}), $k^*$, though a function of $E$,
takes on integer values and is a constant over a
range of $E$. As $E$ increases from $0$ to its
maximum value, $k^*$ decreases from $N$ to $1$, in steps of unity.  
Thus, the entropy curve consists of $N$ linear segments of slope
$\xi(k^*,k^*-1)/\chi$,  $k^*=1,\ldots N$, 
where the segment with larger energy has smaller slope, thereby implying
that the curve is concave. The piecewise linear character of microcanonical
entropy has its origin in the first order poles of the canonical partition 
function, with the slopes being equal to the strength of the poles
\cite{touchette2}.

\begin{figure} 
\includegraphics[width=\columnwidth]{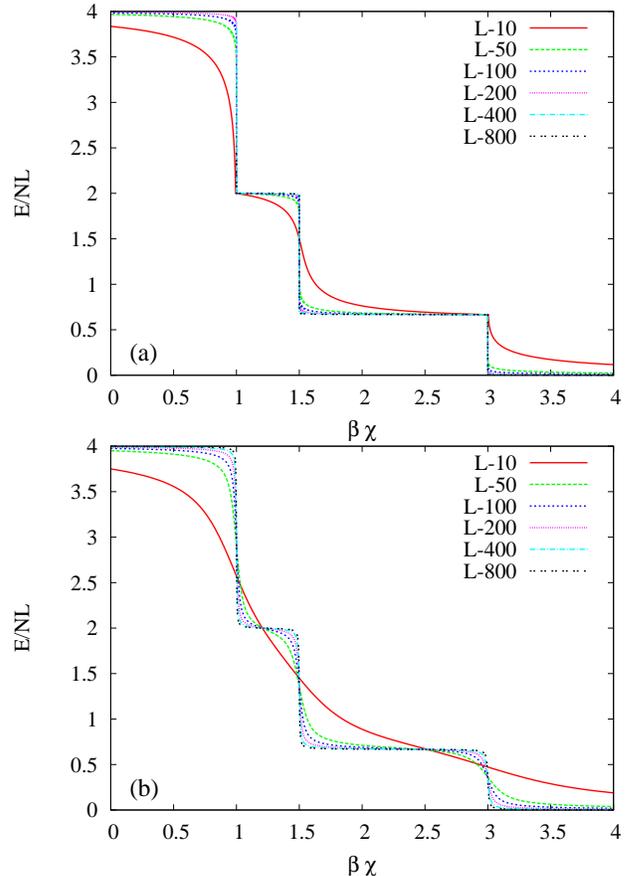}
\caption{The energy--temperature relation for $N=3$, $\chi=3$ for
different $L$ in the (a) microcanonical ensemble and (b) canonical ensemble.
For large values of $L$, the transition
points are given by $(\beta\chi)_c=\xi(k^*,k^*-1)$. 
}
\label{fig2}
\end{figure}

The energy--temperature relation in the microcanonical ensemble
is obtained from the thermodynamic relation 
$\beta=\frac{\partial S}{\partial E}$. This gives 
$\beta\chi=\xi(k^*,k^*-1)$, implying that $\beta \chi$ takes on $N$ distinct
values, corresponding to the $N$ transition points, which coincide
with those obtained from the canonical partition function [see
Eq.~(\ref{eq21})]. In Fig.~\ref{fig2}(a), we show the energy--temperature
relation in the microcanonical ensemble for finite $L$ by considering all the 
terms in $g(E)$.
As $L$ is increased we obtain the limiting step function 
determined by the relation $\beta\chi=\xi(k^*,k^*-1)$. The corresponding
data for the canonical partition function from  Eq.~(\ref{eq14}) 
is shown in Fig.~\ref{fig2}(b).  Thus, in the thermodynamic limit 
($L\rightarrow \infty$), we obtain the same limiting curve in both the
ensembles.

We can further quantify the energy--temperature relation shown in
Fig.~\ref{fig2}.
The energies  of the plateaus are obtained by solving
Eq.~(\ref{eq20}) as an equality, and are given by
$E_{plateau}=\chi N^{-1} (N-k)(N-k+1)L$, $k=0,\ldots, N$. A simple physical
interpretation can be ascribed to these plateaus, as explained below.
Consider a scenario when $m$ counterions have condensed onto the disc, while
the remaining $N-m$ counterions are at the boundary $R$. The energy of this
configuration, in the limit of large $L$ has contribution from two 
parts: (1) $q'^2 L (N-m)^2$ 
corresponding to interaction between a disc of charge $(N-m) q'$ and $N-m$ 
counterions and (2) $q'^2 L (N-m) (N-m-1)/2$ corresponding to interaction
between counterion pairs. The total energy of this geometry is thus $\chi
N^{-1} (N-m) (N-m+1)$. Comparing this energy
with the energy of a plateau, it is clear
that the plateau corresponding to a certain value of $k$ corresponds to a
case where $k$ counterions have condensed and the remaining ones are at
the boundary. 

We stress that we have not used the 
Legendre transformation to evaluate the microcanonical entropy from
the canonical free energy. The entropy obtained by the Legendre
transformation will  
always give the concave envelope of the microcanonical 
entropy~\cite{ellis3,touchette1}. Since the entropy that we have calculated
directly from the density of states is concave, we should be able to 
obtain the same by a Legendre
transform of the free energy, $S(E)=\beta E-\beta F(\beta)$, where $\beta$
is to be eliminated using the energy-temperature relation
$E=\frac{\partial}{\partial\beta}(\beta F)$. In the thermodynamic limit, it
is straightforward to do so for this model.

To summarize, we studied counterion condensation transition on a two
dimensional charged disc in the microcanonical ensemble.
In the limit of infinite dilution, we obtained an expression for the
microcanonical entropy, and showed that the entropy--energy curve
consisted of linear segments with decreasing slope, and hence that the 
entropy is concave with
respect to energy. This implies the equivalence of the microcanonical and
canonical ensembles. In particular, the energy--temperature relation 
and the transition points of CCT obtained from the microcanonical entropy 
are shown to be identical with those obtained from the
canonical partition function. 

\end{document}